\documentclass[Physsubmission, Phys]{SciPost}

\hypersetup{
    colorlinks,
    linkcolor={red!50!black},
    citecolor={blue!50!black},
    urlcolor={blue!80!black}
}

\usepackage[bitstream-charter]{mathdesign}
\DeclareSymbolFont{usualmathcal}{OMS}{cmsy}{m}{n}
\DeclareSymbolFontAlphabet{\mathcal}{usualmathcal}

\newcommand{\ga}{\gamma}
\newcommand{\de}{\delta}

\newcommand{\si}{\sigma}

\newcommand{\ri}{{\mathrm{i}}}

\newcommand{\rR}{{\mathrm{R}}}

\newcommand{\rL}{{\mathrm{L}}}

\newcommand{\rw}{{\mathrm{w}}}

\renewcommand{\L}{{\cal L}}

\newcommand{\ren}{{\mathrm{R}}}

\def\mathswitch#1{\relax\ifmmode#1\else$#1$\fi}
\def\mathswitchr#1{\relax\ifmmode{\mathrm{#1}}\else$\mathrm{#1}$\fi}
\def\mathswitchit#1{\relax\ifmmode{#1}\else$#1$\fi}

\newcommand{\PW}{\mathswitchr W}

\newcommand{\Pf}{f}

\newcommand{\scrs}{\scriptscriptstyle}

\newcommand{\sw}{\mathswitch {s_{\scrs\PW}}}
\newcommand{\cw}{\mathswitch {c_{\scrs\PW}}}

\newcommand{\Qf}{\mathswitch {Q_\Pf}}

\def\refeq#1{\mbox{(\ref{#1})}}

\def\reffi#1{\mbox{Fig.~\ref{#1}}}

\def\citeres#1{\mbox{Refs.~\cite{#1}}}
\def\nn{\nonumber}

\newcommand{\GaBFM}{\hat\Gamma}
\newcommand{\FA}{A}
\newcommand{\FZ}{Z}
\newcommand{\FV}{V}
\newcommand{\FAhat}{\hat A}
\newcommand{\FZhat}{\hat Z}
\newcommand{\FVhat}{\hat V}
\newcommand{\Ff}{f}
\newcommand{\Ffbar}{\bar{f}}

\newcommand*\beq{\begin{equation}}
\newcommand*\eeq{\end{equation}}
\newcommand*\beqar{\begin{eqnarray}}
\newcommand*\eeqar{\end{eqnarray}}
\newcommand*\barr{\begin{array}}
\newcommand*\earr{\end{array}}
\newcommand*\bfi{\begin{figure}}
\newcommand*\efi{\end{figure}}
\newcommand*\btab{\begin{table}}
\newcommand*\etab{\end{table}}
\newcommand*\bpm{\begin{pmatrix}}
\newcommand*\epm{\end{pmatrix}}

\makeatletter
\def\dsl{\mathpalette\make@slash}
\def\make@slash#1#2{\setbox\z@\hbox{$#1#2$}%
  \hbox to 0pt{\hss$#1/$\hss\kern-\wd0}\box0}
\makeatother

\begin{document}

\begin{center}{\Large \textbf{
All-order renormalization of electric charge \\
in the Standard Model and beyond
\\
}}\end{center}

\begin{center}
Stefan Dittmaier
\end{center}

\begin{center}
Albert-Ludwigs-Universit\"at Freiburg,
Physikalisches Institut, 
D-79104 Freiburg, Germany
\\
* stefan.dittmaier@physik.uni-freiburg.de
\end{center}

\begin{center}
\today
\end{center}


\definecolor{palegray}{gray}{0.95}
\begin{center}
\colorbox{palegray}{
  \begin{tabular}{rr}
  \begin{minipage}{0.1\textwidth}
    \includegraphics[width=35mm]{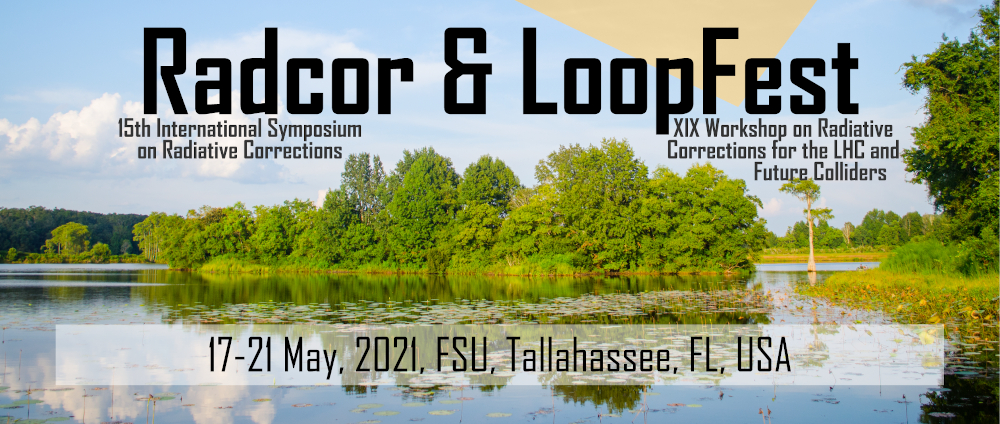}
  \end{minipage}
  &
  \begin{minipage}{0.85\textwidth}
    \begin{center}
    {\it 15th International Symposium on Radiative Corrections: \\Applications of Quantum Field Theory to Phenomenology,}\\
    {\it FSU, Tallahasse, FL, USA, 17-21 May 2021} \\
    \doi{10.21468/SciPostPhysProc.?}\\
    \end{center}
  \end{minipage}
\end{tabular}
}
\end{center}

\section*{Abstract}
{\bf
Electric charge, as defined in the Thomson limit of the
electron--photon interaction vertex, is renormalized
to all orders both in the Standard Model and in any spontaneously
broken gauge theory with gauge group \boldmath{$G\times$}U(1) with a
group factor U(1) that mixes with electromagnetic gauge symmetry.
In the framework of the background-field method the charge 
renormalization constant \boldmath{$Z_e$} is directly obtained from the
photon wave-function renormalization constant, 
similar to the situation in QED, which proves
charge universality as a byproduct.
Exploiting charge universality in arbitrary
\boldmath{$R_\xi$} gauge by formulating the charge renormalization
condition for a ``fake fermion'' that couples only via an
infinitesimal electric charge, \boldmath{$Z_e$} can be expressed in terms of
renormalization constants that are obtained solely from 
gauge-boson self-energies.}

\vspace{10pt}
\noindent\rule{\textwidth}{1pt}
\tableofcontents\thispagestyle{fancy}
\noindent\rule{\textwidth}{1pt}
\vspace{10pt}

\section{Introduction}
\label{sec:intro}

The definition of electric unit charge is carried over from
classical electrodynamics to quantum electrodynamics (QED) 
and to more comprehensive quantum field theories
such as the Standard Model (SM) upon imposing the {\it Thomson renormalization
condition} which demands that the electron--photon interaction vertex for
physical (on-shell) electrons does not receive any radiative
corrections in the limit of low-energy photons.
The charge renormalization constant $Z_e$, which
relates the bare charge $e_0=Z_e e$ to the renormalized unit charge $e$,
can, thus, be determined from
the ee$\gamma$ vertex correction in this limit by direct calculation
of this correction in a given perturbative order.
Exploiting electromagnetic gauge invariance in the form of the
famous Ward identity for the ee$\gamma$ vertex, in QED it is possible to
calculate $Z_e$ from the wave-function renormalization constant
of the photon, i.e.\ merely from a self-energy.
This is not only a welcome technical simplification,
but also an important field-theoretical result that is
helpful in proving structural theorems such as 
{\it charge universality} or 
{\it Thirring's theorem} for low-energy Compton 
scattering~\cite{Thirring:1950cy,Dittmaier:1997dx}.
Charge universality is the non-trivial statement that the
renormalized unit charge does not depend on the
fermion species~$f$ (actually not even on the type of charged particle)
that is used in the formulation of the Thomson
renormalization condition on the $ff\gamma$~vertex.

In the SM and most of its extensions, the derivation of
$Z_e$ from self-energies is very complicated (and not known
beyond the one-loop level) if the derivation is directly
based on gauge invariance, which manifests itself via
Slavnov--Taylor identities for Green functions and Lee identities
for (one-particle irreducible, 1PI) vertex functions.
This is due to the fact that the unbroken electromagnetic U(1)$_{\mathrm{em}}$ 
gauge symmetry is not a mere factor of the gauge group, but
mixes with a non-abelian subgroup in a non-trivial way.
Other U(1) group factors, such as the weak hypercharge group in the
SM, which mix with electromagnetic gauge symmetry cannot
fully play the role of U(1)$_{\mathrm{em}}$ of QED, because they are spontaneously
broken.
Owing to this difficulty, in
the early pioneering proposals for electroweak one-loop 
renormalization~\cite{Ross:1973fp,Sirlin:1980nh,Bardin:1980fe,Fleischer:1980ub,%
Aoki:1982ed,Bohm:1986rj,Hollik:1988ii,Denner:1991kt}
the derivation of $Z_e$ was either based on explicit calculations of
the vertex correction or on Ward identities that were verified by
explicit calculations.
A derivation fully based on Lee identities at one loop without any 
explicit loop calculation has only been given quite 
recently~\cite{Denner:2019vbn};%
\footnote{An alternative proof based on Slavnov--Taylor identities
is sketched in the slides of the corresponding talk given at the
conference.}
owing to its complexity 
a generalization of this approach beyond the one-loop level seems
rather complicated, if not infeasible.

The first all-order result for $Z_e$ in the SM, again expressed in terms of the 
photon wave-function renormalization constant, was given in the
framework of the background-field method (BFM) in Ref.~\cite{Denner:1994xt}. 
This result, in particular, proves charge universality in the SM to
all orders.
In Ref.~\cite{Degrassi:2003rw} it was shown by explicit calculation
that the BFM result for $Z_e$ indeed is in line with the
Thomson condition for the electron--photon vertex at the
two-loop level in arbitrary $R_\xi$ gauge.
Similarly, in 
Refs.~\cite{Actis:2006ra,Actis:2006rb,Actis:2006rc} it was shown 
by explicit two-loop calculation in conventional
't~Hooft--Feynman gauge that the sum
of all genuine vertex corrections and fermionic wave-function corrections to
the $ff\gamma$ vertex vanishes in the Thomson limit;
this is exactly the part in the calculation of $Z_e$ that is ruled by 
gauge invariance but not yet proven on the basis of
Slavnov--Taylor or Lee identities to all orders.
In fact a general result on $Z_e$ expressed in terms of
other renormalization constants that are related to 
gauge-boson self-energies, was suggested
in Ref.~\cite{Bauberger:1997zz}---although correct, this result was
actually more conjectured than derived.%
\footnote{Detailed comments on the arguments given in Ref.~\cite{Bauberger:1997zz}
can be found in the appendix of Ref.~\cite{Dittmaier:2021loa}.}
This form of $Z_e$ was subsequently used in the few existing explicit
electroweak two-loop calculations, such as for the
muon decay~\cite{Freitas:2002ja,Awramik:2002vu}
or Z-boson decay widths~\cite{Dubovyk:2018rlg}.

After briefly sketching the BFM derivation of $Z_e$ within
the SM to all orders, in the following we review the all-order 
derivation~\cite{Dittmaier:2021loa} of
$Z_e$ within arbitrary $R_\xi$ gauge, exploiting charge universality
by imposing the Thomson renormalization condition on the electromagnetic
interaction vertex of a ``fake fermion''~$\eta$ with infinitesimal
electric charge $Q_\eta$ but no other charges or couplings, so that $\eta$
fully decouples from all other particles for $Q_\eta\to0$. 
We first describe the derivation within the SM, where the result
confirms the earlier ``conjecture'' of Ref.~\cite{Bauberger:1997zz},
and then generalize it to a more general gauge group
$G\times$U(1), where $G$ is any compact Lie group and U(1)
has an admixture of electromagnetic gauge symmetry.
This article is just a brief summary of Ref.~\cite{Dittmaier:2021loa},
where a much more complete treatment of the subject 
can be found.

\section{\sloppy Charge renormalization and charge universality in the 
back\-ground-field method}
\label{se:chargerenBFM}

Denoting the relative charge and mass of the fermion~$f$ by $Q_f$ and $m_f$, 
respectively, the Thomson renormalization condition reads
\begin{align}
\label{eq:Affrencond}
\left.\bar{u}(p) \, \GaBFM^{\FAhat\bar ff}_{\ren,\mu}(0,-p,p) \,
u(p)\right\vert_{p^2=m_f^2}
=-Q_f e \,\bar{u}(p)\gamma_\mu u(p),
\end{align}
where $\FAhat$ is the background photon field and
$\GaBFM^{\FAhat\bar ff}_{\ren,\mu}$ the renormalized 
$\FAhat\bar ff$ vertex function in the BFM.
Here $\bar u(p)$ and $u(p)$ are Dirac spinors of the fermion~$f$ with momentum~$p$
fulfilling $p^2=m_f^2$ with the renormalized on-shell mass $m_f$.
In the notation and conventions for all field-theoretical quantities
we follow Ref.~\cite{Denner:2019vbn} throughout.

The needed low-energy limit of the $\FAhat\bar ff$ for on-shell fermions
can be obtained from its BFM Ward identity, which follows from the
background-field gauge invariance of the BFM effective action
(see Refs.~\cite{Denner:1994xt,Denner:2019vbn} and references therein).
This Ward identity for the unrenormalized $\FAhat\bar ff$ 
vertex function reads~\cite{Denner:1994xt}%
\footnote{We ignore fermion generation mixing here; for its restoration see
Ref.~\cite{Dittmaier:2021loa}.} 
\begin{align}
\label{eq:WIAff}
k^\mu \GaBFM^{\FAhat\Ffbar\Ff}_{\mu}(k,\bar p, p) ={} -e_0\Qf 
\left[\GaBFM^{\Ffbar\Ff}(\bar p,-\bar{p}) - \GaBFM^{\Ffbar\Ff}(-p,p)\right],
\end{align}
where $\GaBFM^{\Ffbar\Ff}$ is
the unrenormalized two-point vertex function of the fermions.
To exploit this identity in condition~\refeq{eq:Affrencond},
we have to replace unrenormalized by renormalized quantities.
Indicating bare quantities consistently by subscripts 0,
the relevant parts of this renormalization transformation reads
\begin{align}
f_{0}^\si = \left(Z^{f,\si}\right)^{1/2} \, f^\si, \quad
\bar f_{0}^\si = \left(Z^{f,\si\,*}\right)^{1/2} \, \bar f^\si, 
\quad 
\quad
\bpm \FZhat_{0} \\ \FAhat_{0} \epm ={}
\bpm Z_{\FZhat\FZhat}^{1/2} & Z_{\FZhat\FAhat}^{1/2}  \\[1ex]
     Z_{\FAhat\FZhat}^{1/2} & Z_{\FAhat\FAhat}^{1/2} \epm
\bpm \FZhat \\ \FAhat \epm,
\label{eq:BFMfieldren}
\end{align}
where $\si=\rR,\rL$ refers to the right- and left-handed parts of the
fermion field~$f$ and $\FZhat$ is the background Z-boson field.
The resulting relation between renormalized and unrenormalized
vertex functions reads
\begin{align}
\label{eq:ffrentrafo}
\GaBFM^{\Ffbar\Ff}_{\ren,\mu}(-p,p) ={}&
\left({Z^{f,\si}}^*\right)^{1/2} \,
\left(Z^{f,\si}\right)^{1/2} \,
\GaBFM^{\Ffbar\Ff}_{\mu}(-p,p),
\\
\label{eq:Affrentrafo}
\GaBFM^{\FAhat\Ffbar\Ff}_{\ren,\mu}(k,\bar p, p) ={}&
\sum_{\FVhat=\FAhat,\FZhat} 
Z_{\FVhat\FAhat}^{1/2} \, \left({Z^{f,\si}}^*\right)^{1/2} \,
\left(Z^{f,\si}\right)^{1/2} \,
\GaBFM^{\FVhat\Ffbar\Ff}_{\mu}(k,\bar p, p).
\end{align}
The introduced field-renormalization constants 
$Z^{f,\si}$ for the fermions and $Z_{\hat V\hat V'}$ for the
photon--Z-boson system are fixed by on-shell (OS) 
renormalization conditions, which require canonical normalization
of the residues of particle propagators and eliminates mixing
between different fields for on-shell momenta.
Background-field gauge invariance automatically implies
$Z_{\FZhat\FAhat}=0$
(see, e.g., Refs.~\cite{Denner:1994xt,Denner:2019vbn} for more details).
Using additionally $e_0=Z_e e$,
turns the Ward identity \refeq{eq:WIAff} into an analogous identity for
renormalized quantities,
\begin{align}
\label{eq:WIAffren}
k^\mu \GaBFM^{\FAhat\Ffbar\Ff}_{\ren,\mu}(k,\bar p, p) ={}
-e\Qf \, Z_e \,
Z_{\FAhat\FAhat}^{1/2} \,
\left[\GaBFM^{\Ffbar\Ff}_\ren(\bar p,-\bar{p}) 
- \GaBFM^{\Ffbar\Ff}_\ren(-p,p)\right],
\end{align}
which
is valid for arbitrary momenta $k$, $\bar p$, $p$ obeying $k+\bar p+p=0$.
Taking $k\to0$ for fixed $p$, the terms linear in $k$ obey the relation
\begin{align}
\label{eq:GaBFMAffren}
\GaBFM^{\FAhat\Ffbar\Ff}_{\ren,\mu}(0,-p, p) ={}&
-e\Qf \, Z_e \,
Z_{\FAhat\FAhat}^{1/2} \,
\frac{\partial\GaBFM^{\Ffbar\Ff}_\ren(-p,p)}{\partial p^\mu}. 
\end{align}
Sandwiching this relation between Dirac spinors,
its l.h.s.\ becomes identical to the one of \refeq{eq:GaBFMAffren},
and its r.h.s.\ can be simplified with the OS renormalization
condition for the fermion field, which can be written as
\begin{align}
\label{eq:uGaBFMu}
\bar u(p)\,\frac{\partial\GaBFM^{\Ffbar\Ff}_\ren(-p,p)}{\partial p^\mu}\,u(p) ={}&
\bar u(p)\,\ga_\mu \,u(p).
\end{align}
Thus, combining the charge renormalization condition
\refeq{eq:Affrencond} with \refeq{eq:GaBFMAffren} and \refeq{eq:uGaBFMu}
leads to the simple equation~\cite{Denner:1994xt}
\begin{align}
\label{eq:ZeBFM}
Z_e = Z_{\FAhat\FAhat}^{-1/2}
\end{align}
in the BFM, which is formally identical to the well-known relation in QED.
Note also that Eq.~\refeq{eq:ZeBFM} shows that 
$e_0 \FAhat_{0,\mu}(x) = e \FAhat_\mu(x)$,
i.e.\ that the product of electromagnetic coupling and background
photon field is not renormalized, again in analogy to a QED relation.

\section{\boldmath{Charge renormalization in arbitrary $R_\xi$-gauge}}
\label{se:chargerenrxi}

\begin{figure}
\centerline{
\raisebox{5em}{(a)}
\includegraphics[scale=.9]{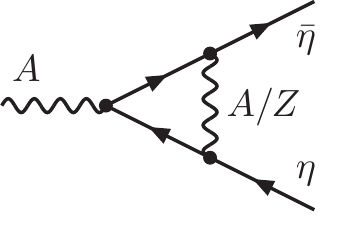} \qquad
\raisebox{5em}{(b)}
\includegraphics[scale=.9]{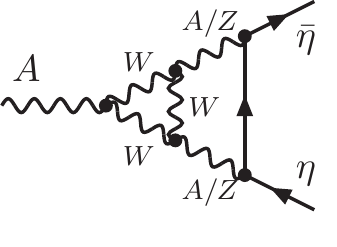} \qquad
\raisebox{5em}{(c)}
\includegraphics[scale=.9]{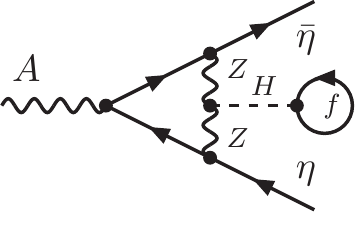}
}
\vspace*{.5em}
\centerline{
\raisebox{4em}{(d)}
\includegraphics[scale=.9]{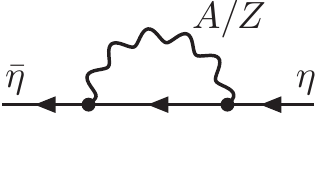} \qquad
\raisebox{4em}{(e)}
\includegraphics[scale=.9]{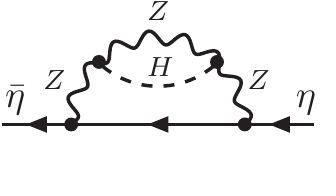} \qquad
\raisebox{4em}{(f)}
\includegraphics[scale=.9]{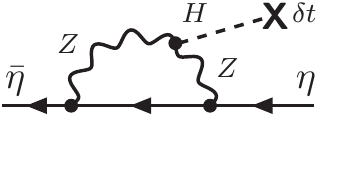}
}
\vspace*{-1.5em}
\caption{Some higher-order diagrams contributing to 
$\Gamma^{\FV\bar\eta\eta}_{\mu}$ (a--c) and 
$\Gamma^{\bar\eta\eta}$ (d--f), which receive
contributions from 1PI diagrams (a, b, d,~e),
from explicit tadpole diagrams (c), and from diagrams
involving tadpole counterterms $\de t$ (f).}
\label{fig:Aetaeta-etaeta-diags}
\end{figure}

We now extend the SM
by adding a fermion field~$\eta$ with vanishing weak isospin, $I_{\rw,\eta}^a=0$, 
and weak hypercharge $Y_{\rw,\eta}=2Q_\eta$.
Taking the limit of vanishing electric charge, $Q_\eta\to0$, the fermion $\eta$ decouples
from all other particles, and we recover the original 
theory---therefore the terminology ``fake fermion''.
The Lagrangian ${\cal L}$ of the SM is, thus, modified by adding
\begin{align}
\label{eq:Leta}
{\cal L}_\eta = 
\bar{\eta} \Bigl( \ri \dsl{\partial} 
- {\textstyle\frac{1}{2}} g_1 Y_{\rw,\eta}\dsl{B} -m_\eta\Bigr) \eta =
\overline{\eta} \left[ \ri \dsl{\partial} 
- Q_\eta e \left(\dsl{A}+\frac{\sw}{\cw}\dsl{Z}\right)
 -m_\eta\right] \eta,
\end{align}
with $m_\eta$ denoting the arbitrary mass of the fermion $\eta$.
We note that the mass term for $\eta$ is gauge invariant, that
$\eta$ is stable,
and that no anomalies are introduced owing to the non-chirality of $\eta$.
As in the SM, $g_1$ is the U(1)$_Y$ gauge coupling, $B^\mu$ the U(1)$_Y$ gauge field,
and $\sw=\sin\theta_\rw$ and $\cw=\cos\theta_\rw$ the sine and cosine 
of the weak mixing angle~$\theta_\rw$.
Employing charge universality, 
we can take the Thomson limit of the $\FA\bar\eta\eta$ vertex
to define the renormalized electric unit charge~$e$,
\begin{align}
\label{eq:Aetaetarencond}
\left.\bar{u}(p) \, \Gamma^{\FA\bar\eta\eta}_{\ren,\mu}(0,-p,p) \,
u(p)\right\vert_{p^2=m_\eta^2}
=-Q_\eta e \,\bar{u}(p)\gamma_\mu u(p).
\end{align}
The relation between $\Gamma^{\FA\bar\eta\eta}_{\ren,\mu}$
and its bare counterpart $\Gamma^{\FA\bar\eta\eta}_{\mu}$ 
follows from the field renormalization transformation 
$\eta_0 = Z^{1/2}_\eta \eta$
and the analog of \refeq{eq:BFMfieldren} for the photon--Z-boson system
and reads
\begin{align}
\label{eq:GaAetaetaren}
\Gamma^{\FA\bar\eta\eta}_{\ren,\mu}(k,\bar p,p) = 
Z_\eta Z_{\FA\FA}^{1/2} \, \Gamma^{\FA\bar\eta\eta}_{\mu}(k,\bar p,p)  
+ Z_\eta Z_{\FZ\FA}^{1/2} \, \Gamma^{\FZ\bar\eta\eta}_{\mu}(k,\bar p,p).
\end{align}
The bare vertex functions $\Gamma^{\FV\bar\eta\eta}_{\mu}$ 
($\FV=\FA,\FZ$) receive lowest-order contributions
and bare vertex corrections $\Lambda^{\FV\bar\eta\eta}_{\mu}$, 
which consist of 1PI loop diagrams and tadpole corrections,
\begin{align}
\Gamma^{\FA\bar\eta\eta}_{\mu}(k,\bar p,p) ={}
- Q_\eta e_0\gamma_\mu +
e_0 \Lambda^{\FA\bar\eta\eta}_{\mu},
\qquad
\Gamma^{\FZ\bar\eta\eta}_{\mu}(k,\bar p,p) ={}
- Q_\eta e_0\frac{s_{\rw,0}}{c_{\rw,0}}\gamma_\mu +
e_0 \Lambda^{\FZ\bar\eta\eta}_{\mu}.
\end{align}
The important observation is now that all diagrammatic contributions
to $\Lambda^{\FV\bar\eta\eta}_{\mu}$ and $Z_\eta$
involve at least
two couplings of photons or Z~bosons to the $\eta$~line that passes through the
whole diagram. 
Some sample diagrams are shown in \reffi{fig:Aetaeta-etaeta-diags}.
For 1PI diagrams it is obvious that at least two couplings to
the $\eta$ line exist, for diagrams with tadpole loops or tadpole counterterms
the same holds true, because
the Higgs field $H$ does not couple to $\eta$.
Since both the photon and the Z~boson couple to $\eta$
proportional to $Q_\eta$, this means that 
$\Lambda^{\FV\bar\eta\eta}_{\mu}={\cal O}(Q_\eta^2)$
and $Z_\eta=1+{\cal O}(Q_\eta^2)$.
Inserting, thus, $\Gamma^{\FA\bar\eta\eta}_{\ren,\mu}$ from
\refeq{eq:GaAetaetaren} into condition \refeq{eq:Aetaetarencond}
and keeping only terms linear in $Q_\eta$ for $Q_\eta\to0$,
we get
\begin{align}
\label{eq:Aetaetarencond2}
-Q_\eta e \,\bar{u}(p)\gamma_\mu u(p) ={}&
\left.\bar{u}(p) \, Z_\eta \left[
Z_{\FA\FA}^{1/2} \, \Gamma^{\FA\bar\eta\eta}_{\mu}(0,-p,p)  
+ Z_{\FZ\FA}^{1/2} \, \Gamma^{\FZ\bar\eta\eta}_{\mu}(0,-p,p)
\right] u(p)\right\vert_{p^2=m_\eta^2}
\nn\\
={}&
- Q_\eta e_0 \left[
Z_{\FA\FA}^{1/2}  + Z_{\FZ\FA}^{1/2} \frac{s_{\rw,0}}{c_{\rw,0}} \right]
\bar{u}(p) \gamma_\mu u(p) \,+\, {\cal O}(Q_\eta^2).
\end{align}
This relation implies
\begin{align}
\label{eq:eren}
e = e_0 \left[
Z_{\FA\FA}^{1/2}  + Z_{\FZ\FA}^{1/2} \frac{s_{\rw,0}}{c_{\rw,0}} \right],
\end{align}
which is the desired relation between $e$ and $e_0$.
Introducing the renormalization constant $\de\cw^2$ according to
\begin{align}
c_{\rw,0}^2  = 1- s_{\rw,0}^2 = \cw^2+\de\cw^2 = 1-\sw^2+\de\cw^2,
\end{align}
we can determine $Z_e$ from \refeq{eq:eren},
\begin{align}
Z_e = \left[
Z_{\FA\FA}^{1/2}  + Z_{\FZ\FA}^{1/2} 
\sqrt{\frac{\sw^2-\de\cw^2}{\cw^2+\de\cw^2}} \right]^{-1}.
\end{align}
This is fully equivalent to the result quoted and used in
\citeres{Bauberger:1997zz,Freitas:2002ja,Awramik:2002vu}.
Since $Z_{\FZ\FA}$ vanishes at tree level,
the renormalization constant $\de\cw^2$ is only 
required to $(\ell-1)$-loop order in the $\ell$-loop calculation of
$Z_e$.

\section{Generalization to non-standard gauge groups}
\label{se:BSM}

The concept of charge universality and charge renormalization outlined above for
the SM can be generalized easily to gauge groups of the type
$G{\times}$U(1), where $G$ is any compact Lie group of rank~$r$ and 
the U(1) group factor plays the analogous
role of weak hypercharge in the SM. 
More precisely, we mean by this that the U(1)$_{\mathrm{em}}$ subgroup of electromagnetic 
gauge symmetry mixes transformations of U(1) and $G$, so that
the photon field $A^\mu$ is a non-trivial linear combination of the 
U(1) gauge field $B^\mu$ and the gauge fields $C_k^\mu$ ($k=1,\dots,r$)
of $G$ corresponding to the diagonal group generators in the Lie algebra of $G$.
For the mechanism of electroweak symmetry breaking 
we only assume that electromagnetic gauge invariance is unbroken.

The original gauge fields $B^\mu$ and $\{C_k^\mu\}$ can be transformed into
fields that correspond to mass eigenstates,
\begin{align}
\begin{pmatrix} B^\mu \\ C_1^\mu \\ \vdots \\ C_r^\mu \end{pmatrix}
= R \begin{pmatrix} A^\mu \\ Z_1^\mu \\ \vdots \\ Z_r^\mu \end{pmatrix},
\qquad
R = \begin{pmatrix} R_{BA} & R_{BZ_1} & \cdots &  R_{BZ_r} \\
R_{C_1A} & R_{C_1Z_1} & \cdots &  R_{C_1Z_r} \\
\vdots & \vdots & \ddots & \vdots \\
R_{C_rA} & R_{C_rZ_1} & \cdots &  R_{C_rZ_r} \end{pmatrix},
\end{align}
where $Z_k^\mu$ ($k=1,\dots,r$) describe neutral massive
gauge bosons similar to the Z~boson of the SM. 
The matrix $R$ is a
generalization of the SM rotation matrix parametrized by the weak mixing angle,
but is not necessarily orthogonal or unitary.
The OS renormalization of the gauge fields proceeds exactly as in the
SM, with an obvious generalization of the $2\times2$ matrix of
renormalization constants $Z_{VV'}$ of \refeq{eq:BFMfieldren}
to a $(r+1)\times(r+1)$ matrix.
The constants $Z_{VV'}$ can be computed from the $VV'$ gauge-boson self-energies
order by order. 

The BFM derivation of the charge renormalization constant generalizes
to the more general gauge group without any problems.
Background-field gauge invariance implies 
$Z_{\FZhat_k\FAhat}=0$ for all $k=1,\dots,r$, i.e.\ there is
no mixing of on-shell photons with any of the Z$_k$ bosons.
The Ward identities \refeq{eq:WIAff} and 
\refeq{eq:WIAffren}, thus, carry over without modification.
As a result, the charge renormalization constant $Z_e$ is
given by $Z_e = Z_{\FAhat\FAhat}^{-1/2}$ as in \refeq{eq:ZeBFM},
proving charge universality as in the SM.

\begin{sloppypar}
To exploit charge universality in the determination of $Z_e$ in arbitrary
$R_\xi$-gauge, we again introduce a fake fermion $\eta$ with the same properties
as above, i.e.\
$\eta$ only carries infinitesimal U(1) charge $Y_{\rw,\eta}=2Q_\eta$, but no 
non-trivial quantum number of $G$, so that the unit charge is given by
$e=g_1 R_{BA}$.
If the model contains 
singlet scalars $S_i$, the scalars $S_i$ may couple to $\eta$ via Yukawa couplings.
The corresponding couplings $y_i$ are free parameters of the model and
can be taken to be
infinitesimally small in analogy to $Q_\eta\to0$, so that decoupling of $\eta$
is guaranteed.
The Lagrangian $\L_\eta$ reads
\begin{align}
\label{eq:LetaBSM}
{\cal L}_\eta = {} &
\bar{\eta} \Bigl( \ri \dsl{\partial} 
- {\textstyle\frac{1}{2}} g_1 Y_{\rw,\eta}\dsl{B} -m_\eta
-\sum_i y_i S_i \Bigr) \eta 
\nn\\
= {} &\overline{\eta} \bigg[ \ri \dsl{\partial} 
- Q_\eta e
\bigg( \dsl{A}+\sum_k \frac{R_{BZ_k}}{R_{BA}} \dsl{Z}_k \bigg) -m_\eta
-\sum_i y_i S_i \bigg] \eta.
\end{align}
Following the same reasoning as for the SM above, the
renormalized $\FA\bar\eta\eta$ vertex function is given by
\begin{align}
\label{eq:GaAetaetarenBSM}
\Gamma^{\FA\bar\eta\eta}_{\ren,\mu}(k,\bar p,p) = 
Z_\eta Z_{\FA\FA}^{1/2} \, \Gamma^{\FA\bar\eta\eta}_{\mu}(k,\bar p,p)  
+ \sum_k Z_\eta Z_{\FZ_k\FA}^{1/2} \, \Gamma^{\FZ_k\bar\eta\eta}_{\mu}(k,\bar p,p),
\end{align}
with the 
unrenormalized $\FV\bar\eta\eta$ vertex functions
\begin{align}
\Gamma^{\FA\bar\eta\eta}_{\mu}(k,\bar p,p) ={}
- Q_\eta e_0\gamma_\mu +
e_0 \Lambda^{\FA\bar\eta\eta}_{\mu},
\qquad
\Gamma^{\FZ_k\bar\eta\eta}_{\mu}(k,\bar p,p) ={}
- Q_\eta e_0\,\frac{R_{0,BZ_k}}{R_{0,BA}}\,\gamma_\mu +
e_0 \Lambda^{\FZ_k\bar\eta\eta}_{\mu}.
\end{align}
Again the vertex corrections $\Lambda^{\FA\bar\eta\eta}_{\mu}$ and
$\Lambda^{\FZ_k\bar\eta\eta}_{\mu}$ as well as the field renormalization constant
$Z_\eta$ receive only corrections that are suppressed at least
by quadratic factors in the new couplings, such as $Q_\eta^2$ or $Q_\eta y_i$.
Typical diagrams contributing to those corrections at the order ${\cal O}(Q_\eta^2)$
(or higher in $Q_\eta$)
can be obtained from the graphs shown in \reffi{fig:Aetaeta-etaeta-diags}
upon interpreting the field $Z$ as any of the $Z_k$ and taking the Higgs
field $H$ as a any Higgs field of the model.
Equation~\refeq{eq:Aetaetarencond2} then generalizes to the considered model
in an obvious way, and we obtain the final result for the 
charge renormalization constant:
\begin{align}
\label{eq:ZeBSM}
Z_e = \left[
Z_{\FA\FA}^{1/2}  + \sum_k Z_{\FZ_k\FA}^{1/2} \,
\frac{R_{BZ_k}+\de R_{BZ_k}}{R_{BA}+\de R_{BA}} \right]^{-1},
\end{align}
where $\de R_{BA}$ and $\de R_{BZ_k}$ are renormalization constants 
for the matrix elements of $R$, i.e.\ $R_0=R+\de R$. 
Recalling that the constants $Z_{\FZ_k\FA}$ vanish at tree level,
we see that the constants $\de R_{BA}$ and $\de R_{BZ_k}$ are only
required to $(\ell-1)$-loop order in the $\ell$-loop calculation of
$Z_e$.
\end{sloppypar}

The case of the SM is trivially recovered from the results of this section
upon identifying
$G={}$SU(2)$_\rw$, $r=1$, $Z_1^\mu=Z^\mu$, $R_{BZ_1}=\sw$, and $R_{BA}=\cw$.

\section{Conclusion}

Employing the property of charge universality, the determination of
the charge renormalization constant $Z_e$ in arbitrary $R_\xi$ gauge
can be greatly simplified upon
applying the Thomson renormalization condition to 
a ``fake fermion'' that has infinitesimal electric charge but no other
charges or couplings. Both in the SM and in the wider class of gauge theories
with gauge group $G\times$U(1), where the U(1) subgroup contains some
component of electromagnetic gauge transformations, $Z_e$ can
be deduced from gauge-boson wave-function and parameter
renormalization constants which can be calculated from gauge-boson 
self-energies only.
Using Slavnov--Taylor or Lee identities to derive this result is
already very cumbersome at the one-loop level, and a consistent
generalization beyond one loop is not known, if not infeasible.

The assumed property of charge universality can for instance proven
in the framework of the background-field method from which it is known
that $Z_e$ can be directly obtained from the photon wave-function
renormalization constant.

For even more general gauge groups without a U(1) factor mixing with
electromagnetic gauge symmetry, the strategy with the 
decoupling ``fake fermion'' seems not to be applicable. 
In this case the consistent use of the background-field method, however,
still bears the possibility to calculate the charge renormalization
constant in terms of gauge-boson wave-function renormalization 
constants in contrast to conventional $R_\xi$-like gauges.

\nolinenumbers

\end{document}